\documentclass[aps,twocolumn]{revtex4}
\usepackage{graphicx}
\usepackage{epsfig}             

\begin{document}

\title{Can optical squeezing be generated via polarization
self-rotation in a thermal vapour cell?}

\author{M.~T.~L.~Hsu, G.~H\'etet, A.~Peng, C.~C.~Harb,
H.-A.~Bachor, M.~T.~Johnsson, J.~J.~Hope and P.~K.~Lam}

\email[Email: ]{ping.lam@anu.edu.au}

\affiliation{Australian Centre for Quantum-Atom Optics, Department of
Physics, Australian National University, ACT 0200, Australia}

\author{A.~Dantan, J.~Cviklinski, A.~Bramati and
M.~Pinard}

\affiliation{Laboratoire Kastler Brossel, Universit\'e Pierre et Marie
Curie, case 74, 75252 Paris cedex 05, France}

\date{\today}

\begin{abstract}
The traversal of an elliptically polarized optical field through a
thermal vapour cell can give rise to a rotation of its polarization
axis.  This process, known as polarization self-rotation (PSR), has
been suggested as a mechanism for producing squeezed light at atomic
transition wavelengths.  In this paper, we show results of the
characterization of PSR in isotopically enhanced Rubidium-87 cells, performed
in two independent laboratories.  We observed that, contrary to
earlier work, the presence of atomic noise in the thermal vapour
overwhelms the observation of squeezing.  We present a theory that
contains atomic noise terms and show that a null result in squeezing
is consistent with this theory.
\end{abstract}
\maketitle

\section{Introduction}

Squeezing is the reduction of the noise variance of an optical field
below the quantum noise limit (QNL).  Many applications, ranging from
increased sensitivity of interferometric measurements \cite{mckenzie}
to quantum entanglement based information protocols \cite{ou,furusawa,
bowen}, are reliant on squeezed light.  Recently, Duan {\it et al.}
\cite{duan} proposed a long-distance quantum communication network
that is based on the interaction of atomic ensembles with squeezed and
entangled light beams.  To achieve such goals, squeezed light at
atomic wavelengths is required.

Conventionally, squeezing can be generated via efficient non-linear
optical processes, such as $\chi^{(2)}$ parametric down-conversion
\cite{ou, lam, sorensen}.  The transparency windows of non-linear
optical crystals, however, may not coincide with some atomic
transitions.  For example, commonly used Sodium and Rubidium atomic transition wavelengths are difficult to access via $\chi^{(2)}$ crystals. Another
method of generating squeezed light is to utilize the $\chi^{(3)}$
atomic Kerr effect at the required atomic wavelength.  These
experiments, however, require ultra-cold atoms confined in cavities
and are therefore technically challenging \cite{lambrecht, josse1}.

Recently, there has been a proposal for generating atomic wavelength
squeezing via the single traversal of an optical field through a
thermal vapour cell \cite{matsko}.  This proposal promises a
simple, scalable and cost-effective means of generating squeezed
light for Rb and potentially for other atomic species.  Due to the
ac Stark shift and optical pumping-induced refractive index changes
of the atomic vapour, an elliptically polarized input field will
experience an intensity dependent rotation of the optical
polarization axes \cite{rochester}. This effect, known as polarization self-rotation (PSR), was suggested as a non-linear mechanism for squeezing \cite{novikova,matsko}. Assuming negligible atomic spontaneous emission noise, Matsko {\it et al.} \cite{matsko} developed a phenomenological model that treats PSR as a cross-phase modulation mechanism.  In the
situation of a linearly polarized input field propagating through
the vapour cell, a non-linear cross-phase interaction occurs between
the two circularly polarized field components.  This results in the
squeezing of the output vacuum field mode that is orthogonally
polarized to the input field.  Analogous to cross-phase modulation
squeezing in optical fibres \cite{haus, margalit, bergman, bergman1,
silberhorn, rosenbluh}, it was suggested that 6 dB of PSR squeezing is possible with thermal Rb vapour cell.  Subsequently Ries {\it et al.}
\cite{ries} reported an observation of 0.85~dB maximum squeezing
from a Rb vapour cell and attributed their squeezing to PSR.

The phenomenological model of PSR squeezing by Matsko {\it
et al.} \cite{matsko} ignored effects such as atomic spontaneous emission. In contrast, Josse {\it et al.} \cite{josse} pointed out the
importance of noise terms arising from the atomic dynamics that could
possibly degrade, if not totally destroy, squeezing.  The model of
Josse {\it et al.} \cite{josse} was based on the interaction of a
linearly polarized field with 4-level atoms.  They showed that in the
high saturation regime, the atomic noise contribution could
potentially be larger than the squeezing term.  Nevertheless, in the
low saturation regime and at sideband frequencies larger than the
atomic relaxation rate, squeezing on the vacuum mode can be generated
via the cross-Kerr effect induced by the bright field.  Such a regime,
however, can only be obtained with ultra-cold trapped atoms enclosed in an
optical cavity \cite{josse1}.

This paper is structured as follows - In Section~\ref{thy}, we review the
theoretical works of Matsko {\it et al.} \cite{matsko} and Josse {\it
et al.} \cite{josse}.  We modified the analysis of Josse {\it et al.}
\cite{josse} to the case of a single traversal optical field through a thermal
vapour cell.  In Section~\ref{expt}, we report measurements of both
the transmittivity and the PSR of an elliptically polarized field
through an isotopically enhanced $^{87}$Rb vapour cell on both the
$D_{1}$ and $D_{2}$ lines.  We then study the noise properties of
the outgoing vacuum field.  The parameter regime investigated extends
beyond the squeezing regime reported in Ref.~\cite{ries}.  In
contradiction to the results in Ref.~\cite{ries}, no optical squeezing was observed.  Instead, we observed excess quadrature noise above the QNL for a wide range of parameters. Finally, in Section~\ref{sec:discussion} we relate experimental
results to the theory and show that under our experimental conditions,
where atomic spontaneous emission is significant, squeezing is
overwhelmed by atomic noise terms.

\section{Theory}
\label{thy}
\subsection{Cross-phase modulation squeezing}
For cross-phase modulation squeezing in fibers, a bright input optical
pulse in the $x$-polarization is delivered into a weakly birefringent
optical fiber.  As a result of the $\chi^{(3)}$ non-linearity in the
fiber, the annihilation ($\hat{a}_{y}$) and creation
($\hat{a}_{y}^{\dagger}$) operators for the $y$-polarized vacuum field
become coupled \cite{haus, margalit, bergman, bergman1}.  The equation
of motion for the $y$-polarized field, in the rotating frame, is given
by
\begin{equation} \label{hauseqn}
\frac{\partial}{\partial z} \hat{a}_{y} (z,t) = i \frac{\kappa}{3} (2
| \langle \hat{a}_{x} \rangle|^2 \hat{a}_{y} + \langle \hat{a}_{x}
\rangle^{2} \hat{a}_{y}^{\dagger} )
\end{equation}
where $\kappa = n_{2} \hbar \omega_{0}^{2} / (c A)$ is the Kerr
coefficient, $n_{2}$ is the non-linear index coefficient of the medium, $\omega_{0}$ is the carrier frequency and $A$ is the effective transverse area of
the propagating field.  The last term of Eq.~(\ref{hauseqn}) describes
the cross-Kerr coupling between the bright $x$- and vacuum
$y$-linearly polarized fields, and is responsible for generating
squeezing in the $y$-polarized field.

Matsko {\it et al.} \cite{matsko} proposed that the PSR
effect in atomic vapour can be used to generate vacuum squeezing.
Their proposal was related to the mechanism of cross-phase coupling
between two orthogonal polarization fields.  We consider the
PSR effect \cite{rochester}, where an elliptically polarized
field undergoes a rotation in its polarization ellipse upon
propagation through an atomic medium.  For an optically thin medium,
the rotation angle is given by
\begin{equation} \label{srang}
\phi = \mathcal{G} \epsilon(0) l
\end{equation}
where $\mathcal{G}$ is the PSR parameter (dependent on the input field
intensity and frequency), $\epsilon(0)$ is the input field ellipticity
(assumed to be small and constant during propagation, $\epsilon(0) =
\epsilon(l)$) and $l$ is the length of the medium.  One could take the
analogy of the PSR effect to the quantum regime by
considering a bright linearly $x$-polarized input field.  The
PSR effect projects fluctuations of the bright $x$-polarized
field onto the $y$-polarized vacuum field.  The relative phase between
the $x$- and $y$-polarized fields then provides amplification or
attenuation of the $y$-polarized field.  This effect could potentially
result in the reduction of the quantum fluctuations of the
$y$-polarized field.

We will now introduce a methodical representation for our optical
field.  For a measurement performed in an exposure time $T$, a freely
propagating single-mode optical field can be described by the electric
field operator given by
\begin{equation}
\hat{E} (z,t) = \mathcal{E}_0 \left( \hat{a} (z,t) e^{i(kz - \omega
t)}
 + \hat{a}^{\dagger} (z,t) e^{-i(kz - \omega t)} \right)
\end{equation}
where $\mathcal{E}_0=\sqrt{\frac{\hbar \omega}{2 \epsilon_{0} cT \mathcal{A}}}$, $\hat{a}(z,t)$ and $\hat{a}^{\dagger}(z,t)$ are the slowly varying
field envelope annihilation and creation operators, respectively.  $z$
is the field propagation axis, $\omega$ is the field carrier frequency
and $\mathcal{A}$ is the quantisation cross-section area.  We can simplify the
expression by introducing $\chi = kz - \omega t$ and
phenomenologically extend the classical PSR to the
quantum regime.  The resulting $y$-polarized field at the output of
the PSR medium is given by
\begin{eqnarray}
\hat{E}_{y}(l) & = & \mathcal{E}_0 \Big[ \hat{a}_{y}(0)
\left( e^{i\chi} - i\mathcal{G}l \cos \chi \right) \nonumber\\
& & + \hat{a}_{y}^{\dagger} (0) \left(e^{-i\chi} + i\mathcal{G}l \cos \chi\right) \Big]
\end{eqnarray}
The noise variance for the $\hat{E}_{y}(l)$ field, taking into account
a phenomenogical absorption parameter $\alpha$ \cite{matsko}, is given
by
\begin{eqnarray}
\langle \hat{E}_{y}^{\dagger}(l) \hat{E}_{y}(l) \rangle & = &
\mathcal{E}_0
\Big[ \Big( 1 - 2\mathcal{G}l \sin \chi \cos \chi \nonumber\\
& & + \mathcal{G}^2 l^2 \cos^2 \chi \Big) e^{-\alpha l} + (1 - e^{-\alpha l} ) \Big] 
\end{eqnarray}
where for appropriate values of the phase $\chi$, squeezing of the
$y$-polarized field can be observed.  Such a model predicts squeezing
values of 6-8~dB below the QNL. However, crucial details such as
spontaneous emission and atomic noise are completely ignored, the
effects of which can reduce, if not completely destroy, squeezing.

\subsection{Squeezing in a 4-level System}\label{sec:theory}

\begin{figure}
\includegraphics[width=\columnwidth]{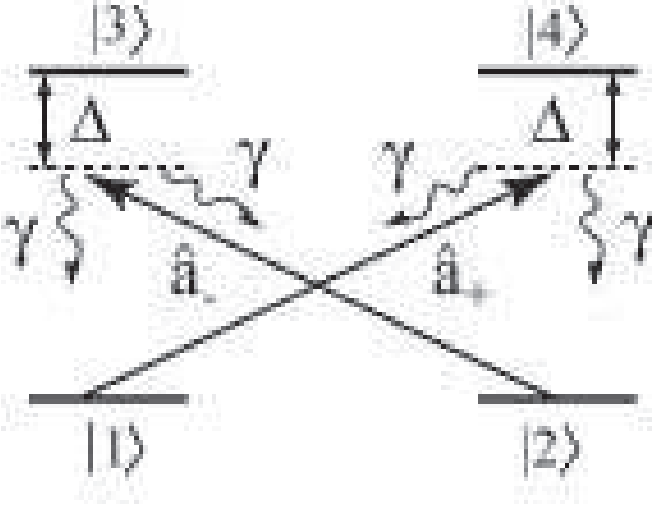}
\caption{Two orthogonal $\sigma_{+}$ and $\sigma_{-}$ circularly
polarized light fields interacting with a 4-level atomic system.}
\label{4level}
\end{figure}

Since optical pumping is the main cause of PSR in the high saturation regime \cite{matsko, zibrov}, which is the relevant regime in our experiment, we can approximate the $D_{1}$ and $D_{2}$ lines of $^{87}$Rb using a 4-level atom model. In such a regime, the influence of atomic coherences are negligible. We thus explore the alternative cross-Kerr squeezing model proposed by Josse {\it et al.} \cite{josse}. In the model, 4-level atoms interact with two orthogonal circularly polarized fields, as shown in Fig.~\ref{4level}. In the experiment of Ref.~\cite{josse1}, squeezing was obtained in the vacuum field (orthogonally polarized to the bright input field) from ultra-cold trapped atoms, enclosed in a cavity. The 4-level squeezing model approximated the level structure of ultra-cold Cesium atoms ($|6S_{1/2}, F=4\rangle$ to $|6P_{3/2}, F=5\rangle$), used in the experiment. In this section, we extend this cavity model to a single-propagation
scenario for a single-mode bright $x$-polarized input
field. We derive the equation of motion describing the noise
fluctuations of the output $y$-polarized vacuum field.
The interaction Hamiltonian is given by
\begin{eqnarray}
\hat{\mathcal{H}}_{\rm int} & = & \hbar n A_{\rm eff} \int_{0}^{l} dz
\Big[ \Delta \hat{\sigma}_{44}(z,t) + \Delta\hat{\sigma}_{33}(z,t)
\nonumber\\
& & - g \Big(\hat{a}_{+}(z,t) \hat{\sigma}_{41}(z,t) +
\hat{a}_{+}^{\dagger}(z,t) \hat{\sigma}_{14}(z,t) \nonumber\\
& & + \hat{a}_{-}(z,t) \hat{\sigma}_{32}(z,t) +
\hat{a}_{-}^{\dagger}(z,t) \hat{\sigma}_{23}(z,t) \Big) \Big]
\end{eqnarray}
where $\hat{a}_{+}(z,t)$ and $\hat{a}_{-}(z,t)$ are the respective
slowly varying field envelope operators for the $\sigma_{+}$ and
$\sigma_{-}$ circularly polarized fields, $n$ is the atomic density
and $g$ is the atom-field coupling constant. The atomic dipole
operator at position $z$ in the rotating frame is defined by locally
averaging over a transverse slice containing many atoms
\begin{equation}
\hat{\sigma}_{ij} (z,t) = \frac{1}{nA \delta z} \sum_{z_{k} \in
\delta z} e^{\frac{i (\omega_{i} -\omega_{j}) z_{k}}{c}} | i
\rangle_{k} \langle j |_{k}
\end{equation}

The optical Bloch equations for the atomic variables
are then given by
\begin{eqnarray}
\frac{\partial}{\partial t} \hat{\sigma}_{14} & = & -(\gamma + i
\Delta) \hat{\sigma}_{14}
+ i g \hat{a}_{+} ( \hat{\sigma}_{11} - \hat{\sigma}_{44} ) +
\hat{F}_{14} \nonumber\\
\frac{\partial}{\partial t} \hat{\sigma}_{23} & = & -(\gamma + i
\Delta) \hat{\sigma}_{23}
+ i g \hat{a}_{-} ( \hat{\sigma}_{22} - \hat{\sigma}_{33} ) +
\hat{F}_{23} \nonumber\\
\frac{\partial}{\partial t} \hat{\sigma}_{11} & = & \gamma
(\hat{\sigma}_{33} + \hat{\sigma}_{44}) -ig\hat{a}_{+}
\hat{\sigma}_{41}+ ig\hat{a}_{+}^{\dagger} \hat{\sigma}_{14} +
\hat{F}_{11} \nonumber\\
\frac{\partial}{\partial t} \hat{\sigma}_{22} &= & \gamma
(\hat{\sigma}_{33} + \hat{\sigma}_{44} ) -ig\hat{a}_{-}
\hat{\sigma}_{32} + ig\hat{a}_{-}^{\dagger} \hat{\sigma}_{23} +
\hat{F}_{22} \nonumber\\
\frac{\partial}{\partial t} \hat{\sigma}_{33} & = & -2\gamma
\hat{\sigma}_{33}
+ig\hat{a}_{-} \hat{\sigma}_{32} - ig\hat{a}_{-}^{\dagger}
\hat{\sigma}_{23} + \hat{F}_{33} \nonumber\\
\frac{\partial}{\partial t} \hat{\sigma}_{44} & = & -2\gamma
\hat{\sigma}_{44} +ig\hat{a}_{+} \hat{\sigma}_{41} -
ig\hat{a}_{+}^{\dagger} \hat{\sigma}_{14} + \hat{F}_{44}
\nonumber\\\label{eqatoms}
\end{eqnarray}
where we have introduced the spontaneous decay term
$\gamma$ and Langevin noise operators $\hat{F}_{ij}$ that arise from
the coupling of atoms to a vacuum reservoir. The Maxwell wave equations describing the $\sigma_{+}$ and $\sigma_{-}$-polarized optical fields are given respectively by
\begin{eqnarray}
\left( \frac{\partial}{\partial t} + c \frac{\partial}{\partial z}
\right)
\hat{a}_{+} (z,t) & = & igN \hat{\sigma}_{14} (z,t) \label{eqA+}\\
\left( \frac{\partial}{\partial t} + c \frac{\partial}{\partial z}
\right) \hat{a}_{-} (z,t) & = & igN \hat{\sigma}_{23} (z,t)
\label{eqA-}
\end{eqnarray}
where $N$ is the total number of atoms. To deduce the noise properties of the field, we linearize the equations around the semi-classical steady state, and write the operators in the form $\hat{a}=\langle \hat{a}\rangle+\delta
\hat{a}$. Transforming into the Fourier domain and linearizing
Eqs.~(\ref{eqatoms})-(\ref{eqA-}) yields the equation of motion for
the quantum fluctuations of the $y$-polarized vacuum mode
$\hat{a}_{y} = -i(\hat{a}_{+} + \hat{a}_{-})/\sqrt{2}$, given by
\begin{equation} \label{sqz}
\frac{\partial}{\partial \bar{z}} \delta\hat{a}_{y}= -\Gamma(\omega)
\delta\hat{a}_{y} + \kappa(\omega) \Big(\delta\hat{a}_{y}
-\delta\hat{a}_{y}^{\dagger}\Big)  + \hat{F}_{y}
\end{equation}
where $\bar{z}=z/l$ and
\begin{eqnarray} \label{eqkappaomega}
\kappa(\omega)&=&\kappa(0)\Lambda(\omega)\\
\Gamma(\omega)&=&-i\omega\frac{l}{c}+\kappa(\omega)+\kappa(0)^*\Lambda'(\omega)
\nonumber\\
\kappa(0)&=&\frac{C\gamma}{2(\gamma+i\Delta)}\frac{1}{1+s} \nonumber\\
\Lambda(\omega)&=&\frac{I_x(\gamma-i\omega)(2\gamma-i\omega)}
{2I_x(\gamma-i\omega)^2-i\omega(2\gamma-i\omega)[(\gamma-i\omega)^2+\Delta^2]}
\nonumber \\
\Lambda'(\omega)&=&i\omega\frac{I_x(\gamma-i\omega)-(\gamma-i\Delta)(\gamma-i\Delta-
i\omega)(2\gamma-i\omega)}
{2I_x(\gamma-i\omega)^2-i\omega(2\gamma-i\omega)[(\gamma-i\omega)^2+\Delta^2]}
\nonumber
\end{eqnarray}
where $C=g^2Nl/\gamma c$ is the cooperativity parameter,
$I_{x}=|g\langle \hat{a}_{x}\rangle|^2$ is the mean field intensity
and $s= I_{x}/(\gamma^2+\Delta^2)$ is the saturation parameter. The
last term of Eq.~(\ref{sqz}) represents the atomic Langevin noise
term and is responsible for a loss or degradation of squeezing. Its
exact form and noise spectrum are given and discussed in Sec.
\ref{sec:discussion}.

Note that for $\omega=0$, the imaginary part of $\kappa(0)$ from the second term on the right hand side of Eq.~(\ref{sqz}) equates to the first term on the right hand side of Eq.~(\ref{hauseqn}). This turns out to be the parameter $\mathcal{G}l$ given in Eq.~(\ref{srang}). In the 4-level atom model, the PSR for one velocity class increases with the number of atoms and is maximum when $\Delta^2=\gamma^2+I_{x}$. For a Doppler-broadened vapour, $\mathcal{G}l$ can be obtained by summing Eq.~(\ref{eqkappaomega}) over all the velocity classes. Note that $\kappa(\omega)$ also gives the amplitude of the cross-Kerr squeezing term in $\delta \hat{a}_y^{\dagger}$, as in Eq.~(\ref{sqz}). However, the associated atomic noise contribution must be evaluated in order to obtain the
total noise spectrum for the output $y$-polarized field.


\section{Experiment}
\label{expt}
\begin{figure}[!ht]
\begin{center}
\includegraphics[width=\columnwidth]{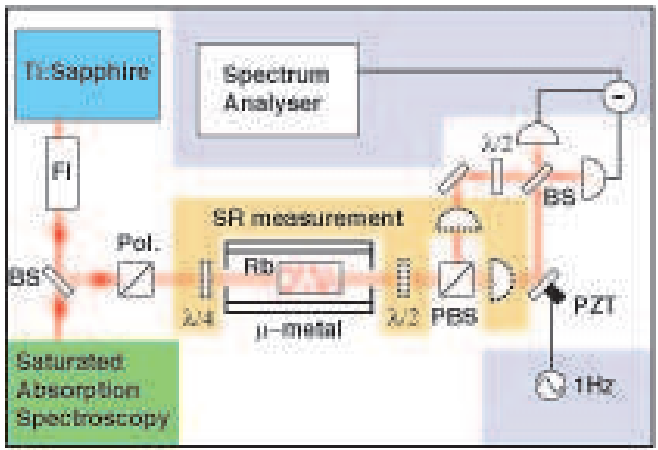}
\caption{Schematic of experimental setup. All polarising optics are of the Glan-Thompson type. FI: Faraday isolator, BS: beam-splitter. Pol.: Polarizer, PBS: polarising beam-splitter, $\lambda/4$: quarter wave-plate, $\lambda/2$: half wave-plate, PZT: piezo-electric actuator.} \label{layout}
\end{center}
\end{figure}

In this section, we present experimental results obtained from the two authoring institutions. Both experiments have similar experimental arrangement. In our experiments, a coherent beam at 795~nm (or 780~nm) was delivered from a Ti:Sapphire laser (Coherent MBR-110), as shown in Fig.~\ref{layout}. The laser beam was measured to be quantum noise limited at sideband frequencies $\geq 1$~MHz. A small fraction of the beam was sent through another Rubidium (Rb) vapour cell for saturated absorption spectroscopy. This provided us with a fine frequency reference for the laser and also allowed the possibility of laser frequency stabilisation. The majority of the beam was sent through a polarizer which transmitted the $x$-polarized field.
In order to measure the PSR and absorption of an input
elliptically polarized beam through the vapour cell, the orange-shaded
configuration of Fig.~\ref{layout} was used.  The $x$-linearly
polarized beam was converted into an elliptically polarized beam using
a $\lambda/4$ wave-plate.  The beam (collimated to a waist size of
$\sim 425 \mu$m) then passed through an isotopically
enhanced $^{87}$Rb vapour cell (75mm length), which was temperature
stabilised at $72^{\circ}$C (which corresponded to an atomic density
of $10^{11}$ atoms/cm$^{3}$).  The vapour cell was enclosed in a
two-layer $\mu$-metal alloy cylinder, with end caps.  The stray
magnetic fields within the shielding region were measured to be $<
2$~mG in all three spatial axes.  The output beam from the cell was
then analysed using a balanced polarimeter setup, which consisted of a
$\lambda/2$ wave-plate, a polarising beam-splitter and two balanced
photo-detectors.  The $\lambda/2$ wave-plate was adjusted to balance
the powers in the $x$- and $y$-linearly polarized beams from the
outputs of the polarising beam-splitter, when the
frequency of the laser was tuned far off-resonance.  Thus any rotation
of the axis of the input elliptically polarized beam could be measured
using the relationship \cite{rochester}
\begin{equation}
\phi = \frac{V_{1} - V_{2}}{2(V_{1} + V_{2})}
\end{equation}
where $V_{1}$ and $V_{2}$ are the DC signals from the photo-detectors.

To measure the quadrature noise properties of the $y$-linearly
polarized vacuum beam, we then performed homodyne detection, as
shown in Fig.~\ref{layout}, using the $x$-linearly polarized output
of the polarising beam-splitter as a local oscillator.

\subsection{Classical results}
\begin{figure}[!ht]
\begin{center}
\includegraphics[width=\columnwidth]{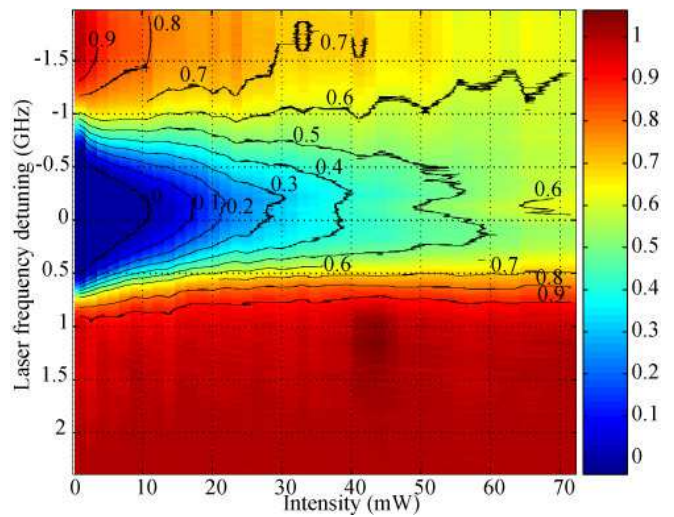}
\caption{False colour contour plot of the normalised transmission
results for the $D_{2}$ line, as a function of input beam intensity
and laser frequency detuning. Zero frequency corresponds to the
$|5^{2}S_{1/2}, F_{g}=2 \rangle$ to $|5^{2}P_{3/2}, F_{e}=3 \rangle$
energy levels.} \label{780abs}
\end{center}
\end{figure}

\begin{figure}[!ht]
\begin{center}
\includegraphics[width=\columnwidth]{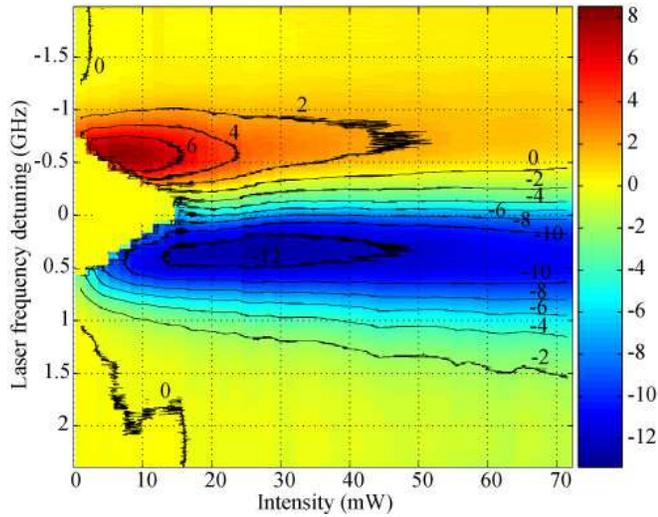}
\caption{False colour contour plot of $\mathcal{G}l$ for the $D_{2}$
line, normalised to the input beam ellipticity of $2^{\circ}$, as a
function of input beam intensity and laser frequency detuning. Zero
frequency corresponds to the $|5^{2}S_{1/2}, F_{g}=2 \rangle$ to
$|5^{2}P_{3/2}, F_{e}=3 \rangle$ energy levels.} \label{780sr}
\end{center}
\end{figure}

\begin{figure}[!ht]
\begin{center}
\includegraphics[width=\columnwidth]{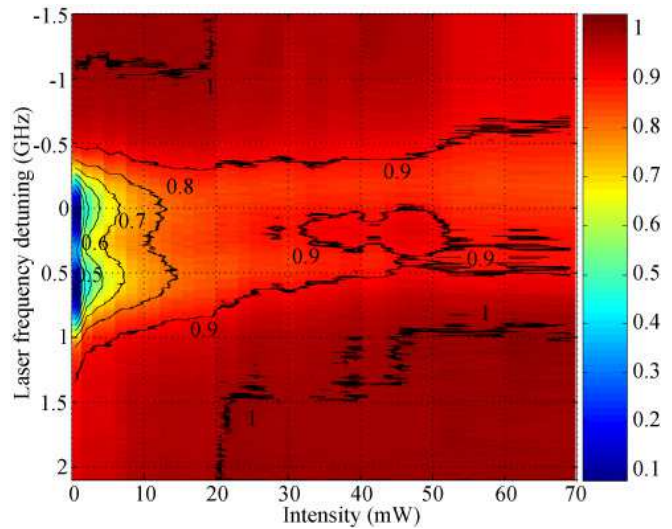}
\caption{False colour contour plot of the normalised transmission
results for the $D_{1}$ line, as a function of input beam intensity
and laser frequency detuning. Zero frequency corresponds to the
$|5^{2}S_{1/2}, F_{g}=2 \rangle$ to $|5^{2}P_{1/2}, F_{e}=1 \rangle$
energy levels.} \label{795abs}
\end{center}
\end{figure}

\begin{figure}[!ht]
\begin{center}
\includegraphics[width=\columnwidth]{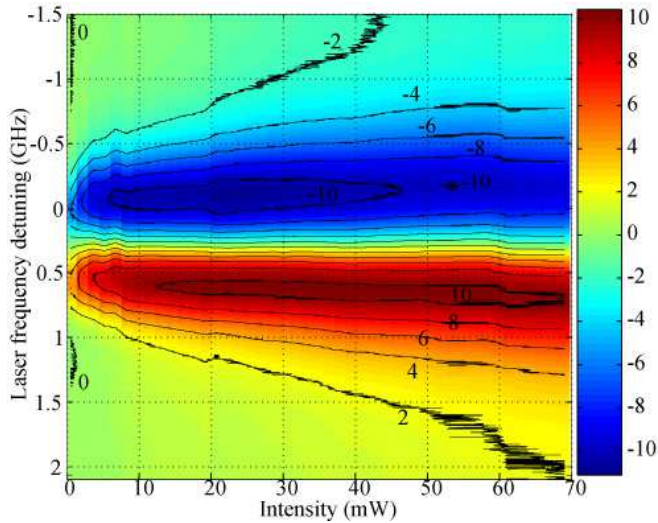}
\caption{False colour contour plot of $\mathcal{G}l$ for the $D_{1}$
line, normalised to the input beam ellipticity of $2^{\circ}$, as a
function of input beam intensity and laser frequency detuning. Zero
frequency corresponds to the $|5^{2}S_{1/2}, F_{g}=2 \rangle$ to
$|5^{2}P_{1/2}, F_{e}=1 \rangle$ energy levels.} \label{795sr}
\end{center}
\end{figure}

The PSR and transmission of an input elliptically
polarized beam through the Rb vapour cell were measured by scanning
the laser frequency across the energy levels of interest, for a
fixed input beam intensity. For the $D_{2}$ line, the relevant levels
were $|5^{2}S_{1/2}, F_{g}=2 \rangle$ to $|5^{2}P_{3/2}, F_{e}=1,2,3
\rangle$ and for the $D_{1}$ line, $|5^{2}S_{1/2}, F_{g}=2 \rangle$ to
$|5^{2}P_{1/2}, F_{e}=1,2 \rangle$. We repeated the measurement for
varying input beam powers and obtained a contour map of
PSR and transmission as a function of laser frequency
detuning and input beam intensity, shown in Figs.~\ref{780abs},
\ref{780sr}, \ref{795abs} and \ref{795sr}.

The transmission results for the $D_{2}$ line are shown in
Fig.~\ref{780abs}.
The region of lowest transmission $<10$~\% occurred at input beam
intensities $\le 15$~mW, around laser frequencies close to zero
detuning. For input beam powers $\ge 30$~mW greater transmission
($\ge 30$~\%) was observed. However, power broadening effects were
also observed for higher input beam powers, with reduced
transmission at frequencies $\le -1$~GHz. The transmission was
non-symmetric with high transmission ($>90$~\%) for frequencies $\ge
1$~GHz, whilst reduced transmission ($>60$~\%) for frequencies $\le
-1$~GHz. This was due to the level structure of the excited states of
the $D_{2}$ line, where the separations between the hyperfine levels
are small (within a frequency band of $\sim 0.5$~GHz). Power
broadening effects were also observed for input beam intensities $\ge
30$~mW.

The PSR results for the $D_{2}$ line are shown in Fig.~\ref{780sr}.
The regions of largest PSR were 0.3~GHz and -0.6~GHz. The
input beam powers which gave the largest $\mathcal{G}l$ magnitudes of
8 and 13 were $\sim 8$~mW and $\sim 30$~mW, respectively. Zero
$\mathcal{G}l$ around zero detuning for input beam powers $\le 15$~mW was due to the low transmission of the input beam for the optically thick $^{87}$Rb vapour cloud. However, at frequency detunings $\ge 0.5$ and $\le -0.5$~GHz, significant
PSR was observed even though the transmission was reduced.
For input beam intensities $\ge 20$~mW, the PSR was
preferentially larger with positive frequency detunings as opposed
to negative frequency detunings. 

\begin{figure}[!ht]
\begin{center}
\includegraphics[width=\columnwidth]{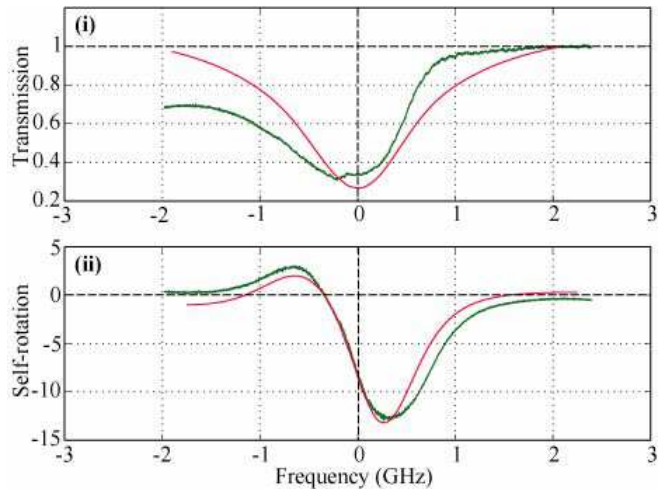}
\caption{The normalised transmission and $\mathcal{G}l$ results for
the $D_{2}$ line are shown in Figures~(i) and (ii), respectively. The
red curves are the theoretical fits to the experimental results
(green curve). Input beam intensity= 31.5~mW, and zero frequency
corresponds to the $|5^{2}S_{1/2}, F_{g}=2 \rangle$ to
$|5^{2}P_{3/2}, F_{e}=3 \rangle$ energy levels.} \label{d2fit}
\end{center}
\end{figure}

In order to explain the asymmetry present in the PSR results, we modelled the hyperfine energy levels of the $D_{2}$ line and took into account Doppler broadening. The theoretical fits to the experimental data are shown in
Fig.~\ref{d2fit}.
The reduction in PSR in the negative frequency detuning
region was due to reduced transmission, as observed in
Fig.~\ref{780abs}. Broadening of the PSR profile was
observed for higher input beam powers.

The transmission results for the $D_{1}$ line are shown in
Fig.~\ref{795abs}.
The region of lowest transmission ($<50$~\%) occured for input beam
intensities $\le 3$~mW. These regions were confined around two
frequency detuning bands, the $-0.2$ to 0.25~GHz band and the 0.4 to
0.8~GHz band. The two frequency bands corresponded to the absorption
lines centred at the $|5^{2}S_{1/2}, F_{g}=2 \rangle$ to
$|5^{2}P_{1/2}, F_{e}=1 \rangle$ and $|5^{2}S_{1/2}, F_{g}=2
\rangle$ to $|5^{2}P_{1/2}, F_{e}=2 \rangle$ energy levels,
respectively. For input beam powers $\ge 5$~mW, significant
transmission was observed ($>70$~\%). For most input beam powers,
the transmission of the $D_{1}$ line was significantly higher than
that of the $D_{2}$ line. This was due to the weaker atom-field
coupling in the $D_{1}$ line compared to the $D_2$ line.

\begin{figure}[!ht]
\begin{center}
\includegraphics[width=\columnwidth]{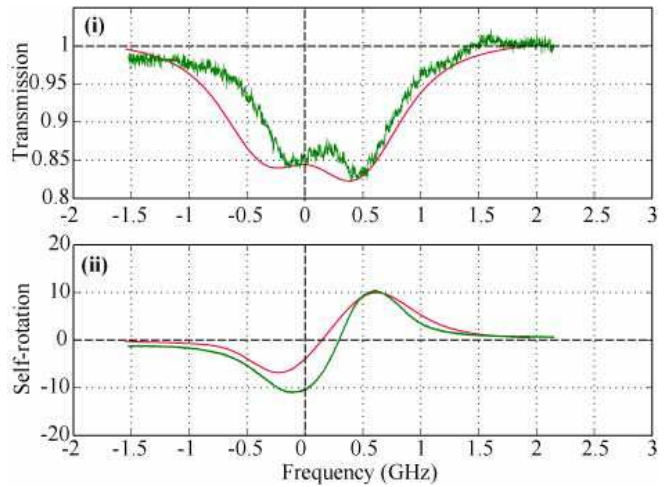}
\caption{The normalised transmission and $\mathcal{G}l$ results for
the $D_{1}$ line are shown in Figures~(i) and (ii), respectively.
The red curves are the theoretical fits to the experimental results
(green curve). Input beam intensity= 22.3~mW, and zero frequency
corresponds to the $|5^{2}S_{1/2}, F_{g}=2 \rangle$ to
$|5^{2}P_{1/2}, F_{e}=1 \rangle$ energy levels.} \label{d1fit}
\end{center}
\end{figure}

The PSR results for the $D_{1}$ line are shown in
Fig.~\ref{795sr}.
The regions of largest PSR occurred at frequency
detunings -0.15~GHz and 0.6~GHz. The input beam powers that gave
the largest $\mathcal{G}l$ magnitudes of 10 and 11 were $\sim 35$~mW
and $\sim 22$~mW, respectively. Significant PSR was
observed for input beam powers $>3$~mW since the transmission was
always $>50$~\%. The $\mathcal{G}l$ magnitude was almost equal in
both frequency bands corresponding to the two absorption lines
centred at the $|5^{2}S_{1/2}, F_{g}=2 \rangle$ to $|5^{2}P_{1/2},
F_{e}=1 \rangle$ and $|5^{2}S_{1/2}, F_{g}=2 \rangle$ to
$|5^{2}P_{1/2}, F_{e}=2 \rangle$ energy levels, for most input beam
powers. This was due to the excited state level structure of the
$D_{1}$ line, where the two excited state levels have a large
separation of $\sim 0.8$~GHz. This is illustrated by modelling the
hyperfine excited state level structure of the $D_{1}$ line. The
theoretical fits to the experimental data are shown in
Fig.~\ref{d1fit}.
The two transmission dips are of approximately the same magnitude,
resulting in the two PSR peaks to be of equal magnitudes.

\subsection{Quantum results}

\begin{figure}[!ht]
\begin{center}
\includegraphics[width=\columnwidth]{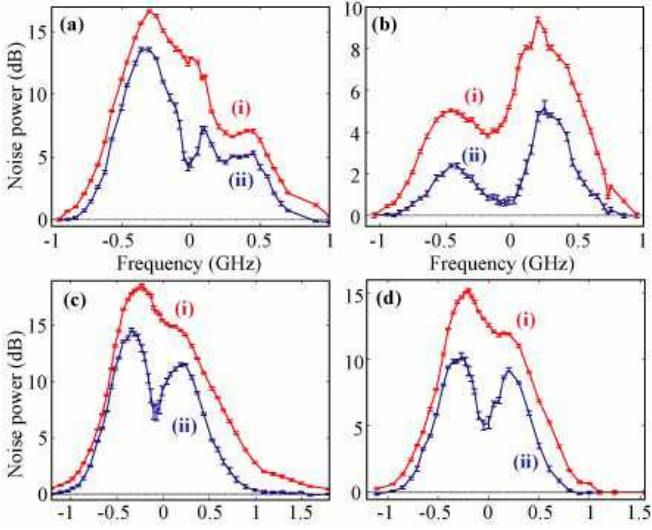}
\caption{(i) Amplitude and (ii) phase quadrature noise results for the $D_2$ line, normalised to the QNL and dark noise-subtracted. Figures
(a) and (b) correspond to an input beam power of 21~mW at sideband
frequencies of 3~MHz and 6~MHz, respectively. Figures (c) and (d)
are results for an input beam power of 35~mW, at sideband
frequencies of 3~MHz and 6~MHz, respectively. Zero frequency
corresponds to the $|5^{2}S_{1/2}, F_{g}=2 \rangle$ to
$|5^{2}P_{3/2}, F_{e}=3 \rangle$ energy level. ResBW: 100~kHz and
VBW: 30~Hz.} \label{780noise}
\end{center}
\end{figure}

\begin{figure}[!ht]
\begin{center}
\includegraphics[width=\columnwidth]{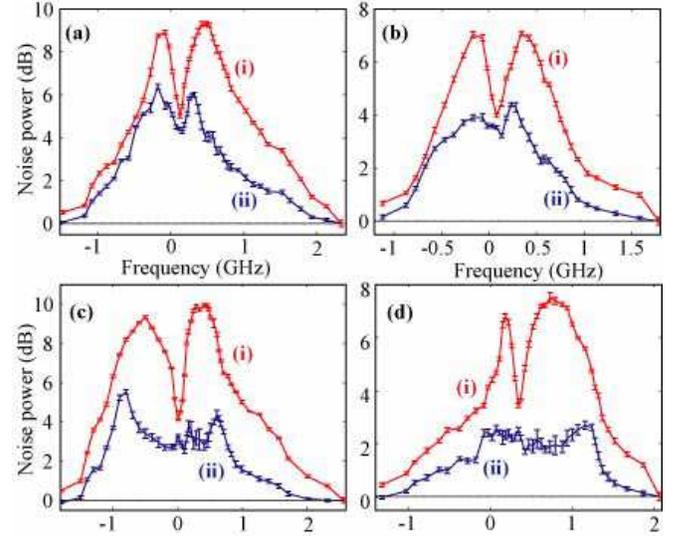}
\caption{(i) Amplitude and (ii) phase quadrature noise results for the D1 line, normalised to the QNL and subtracted by the dark noise. Figures
(a) and (b) correspond to an input beam power of 21~mW at sideband
frequencies of 3~MHz and 6~MHz, respectively. Figures (c) and (d)
are results for an input beam power of 35~mW, at sideband
frequencies of 3~MHz and 6~MHz, respectively. Zero frequency
corresponds to the $|5^{2}S_{1/2}, F_{g}=2 \rangle$ to
$|5^{2}P_{1/2}, F_{e}=1 \rangle$ energy level. ResBW: 100~kHz and
VBW: 30~Hz.} \label{795noise}
\end{center}
\end{figure}

The input field was linearly polarized in the $x$-axis and we
measured the quadrature noise of the outgoing $y$-polarized vacuum
field, using the homodyne detection setup shown in
Fig.~\ref{layout}. The bright $x$-polarized output field was used as
a local oscillator. The fringe visibility of the
interferometer was 99~\%. The two outputs of the interferometer were
then detected using two balanced Silicon photo-detectors (which
consisted of Hamamatsu S3883 photo-diodes with measured quantum
efficiency values of 94.6\%) with bandwidths of $\sim 20$MHz.
Blocking the weak field provided a measurement of the QNL.
The QNL was checked for linearity with beam power and the common mode
rejection was optimised to $\sim 30$-40~dB from 100~kHz to 10~MHz.
We also checked that the polarising beam-splitter was well aligned such that negligible amounts of the $x$-polarized field emerged at the $y$-polarized output port. The result of the noise measurement for various sideband frequencies at various laser frequency detunings and input beam powers, are shown in Figs.~\ref{780noise} ($D_2$ line) and \ref{795noise} ($D_1$ line). 

\begin{figure}[!ht]
\begin{center}
\includegraphics[width=\columnwidth]{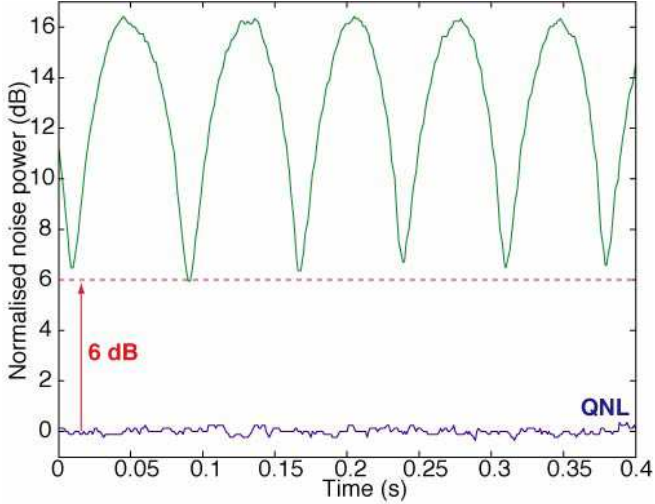}
\caption{Scanned quadrature noise for the $D_2$ line measured in
zero span at a sideband frequency of 3~MHz. The input beam power was
35~mW and the laser frequency was -70~MHz from the $|5^{2}S_{1/2},
F_{g}=2 \rangle$ to $|5^{2}P_{1/2}, F_{e}=3 \rangle$ energy level.
All plots are dark noise subtracted. ResBW: 100~kHz and VBW: 30~Hz.}
\label{noise}
\end{center}
\end{figure}

The largest quadrature noise observed for the $D_{2}$ line
was 10~dB at a detuning of -70~MHz as shown in Fig.~\ref{780noise}~(c). A time scanned quadrature noise measurement is shown in Fig.~\ref{noise}.
In the noise plots of Figs.~\ref{780noise}~(a), (c) and (d), we
observed large levels of excess noise of typically 5~dB above the QNL. In Fig.~\ref{780noise}~(b) the excess noise
level was 0.8~dB above the QNL. This was the lowest noise level
observed around zero detuning. The largest values of the phase
quadrature noise level corresponded to the regions of maximum
PSR as shown in Fig.~\ref{780sr}. At large frequency
detunings from resonance, both quadrature noise levels were reduced to the
QNL.

The noise measurements of the output vacuum field, for the D1 line,
are shown in Fig.~\ref{795noise}.
The largest noise modulation observed was 7~dB which occurred at a
frequency detuning of 150~MHz as shown in Fig.~\ref{795noise}~(c). In
Figs.~\ref{795noise}~(a)-(d), the phase quadrature noise level
around zero detuning was always above the QNL due to the presence of
large excess noise (3-4~dB). The largest values of the amplitude
noise level corresponded to the regions of maximum PSR as
shown in Fig.~\ref{795sr}. At large frequency detunings from
resonance, both quadrature noise levels were reduced to the QNL.

The noise measurement results presented do not vary qualitatively with varying
beam focussing geometry, incident power or temperature. A large
amount of excess noise was systematically observed close to
resonance. We also performed similar experiments using
paraffin-coated cells and cells containing buffer gas, none of which
resulted in the observation of squeezing. Although the PSR and
transmission results measured were in a very similar regime to that
of Ref.~\cite{ries}, the quantum noise results are not in agreement
with either the predictions of Ref.~\cite{matsko} or the observations of
Ref.~\cite{ries}. We use the model presented in Sec.~\ref{sec:theory} to discuss our experimental observations.

\section{Discussion and Conclusions}\label{sec:discussion}

\subsection{Langevin noise analysis}
In order to contrast the effect of the atomic noise terms with the
squeezing term in Eq.~(\ref{sqz}), we consider the Langevin term given
by
\begin{equation}
\hat{F}_{y} = \frac{gNl}{c}\left[(A+B) \hat{f}_{y}+B
\hat{f}_{y}^{\dagger} + i\sqrt{I_{x}}
A\left(\frac{\hat{f}_{z}}{-i\omega}+\frac{\hat{f}_{z}'}{2\gamma-i\omega}\right)\right]
\end{equation}
where
\begin{eqnarray}
A & = &\frac{(\gamma-i\Delta-i\omega)(-i\omega)(2\gamma-i\omega)}{D}
\nonumber\\
B & = & \frac{I_{x}(\gamma-i\omega)}{D} \nonumber\\
D &= &
2I_{x}(\gamma-i\omega)^2-i\omega(2\gamma-i\omega)[(\gamma-i\omega)^2+\Delta^2]
\nonumber
\end{eqnarray}
with $I_{x} = |g \langle \hat{a}_{x} \rangle |^{2}$, $\hat{f}_{y}=(
\hat{F}_{14}+ \hat{F}_{23})/\sqrt{2}$, $\hat{f}_{z} = (\hat{F}_{22}
- \hat{F}_{11})/\sqrt{2}$ and $\hat{f}_{z}' = (\hat{F}_{44} -
\hat{F}_{33})/\sqrt{2}$. The contribution of this noise term, which
depends on the sideband frequency, is to be compared with the
cross-Kerr squeezing term $\kappa(\omega)$. As shown in
Ref.~\cite{josse}, in the low saturation regime, large excess atomic
noise associated with optical pumping on the $y$-polarized field
dominates at sideband frequencies \textit{lower} than the
spontaneous emission rate ($\omega\ll\gamma$). In the low sideband
frequency regime (assuming $\Delta\gg\gamma$), one obtains
\begin{eqnarray}
\frac{\partial}{\partial \bar{z}}\delta \hat{a}_y & = &
i\frac{\delta_{0}}{1+s} \delta\hat{a}_{y}^{\dagger} +
\frac{gNl}{2\gamma c}(\hat{f}_{y} + \hat{f}_{y}^{\dagger} +
\frac{\Delta}{2\sqrt{I_{x}}} \hat{f}_{z})
\end{eqnarray}
where $\delta_{0} = C \gamma/2\Delta$ denotes the linear dephasing.
Ignoring depletion of the mean $x$-polarized field, the Langevin
noise contribution is shown to be proportional to $C/gl$ at least.
For the experiment, this quantity is greater than the QNL, such that large excess noise is present in all quadratures for low sideband frequencies, even when absorption is ignored. One therefore cannot observe squeezing in this regime.

In the experiment, the quantum noise of the vacuum field was measured
only for sideband frequencies \textit{greater} than the excited
state decay rate ($\omega\geq\gamma$).  In this high sideband
frequency regime (assuming $\Delta\gg\gamma$), we obtain succinct
expressions for $\kappa(\omega)$, $\Gamma(\omega)$ and $\tilde{F}_y$ given by
\begin{eqnarray}
\kappa(\omega)&=&\frac{-i\delta_0s}{(1+s)(1+2s)},\hspace{1cm}
\Gamma(\omega)=\frac{-i\delta_0}{1+s}\\ \nonumber \hat{F}_y&\simeq &
-i\frac{gNl}{\Delta c(1+2s)}\big[(1+I_x/\omega\Delta) \hat{f}_{y} +
(I_{x}
/\omega\Delta) \hat{f}_{y}^{\dagger}\\
&&\hspace{3cm}-(\sqrt{I_{x}}/\omega)(\hat{f}_{z} + \hat{f}_{z}')\big].
\end{eqnarray}
The above equations describe the atomic noise contribution that may
degrade the squeezing of the output $y$-polarized vacuum field.

The optimization of squeezing is dependent on finding a regime that
has low absorption and strong non-linearity. We now proceed by
dividing the discussion into low and high atomic transition
saturation regimes.

\subsection{Low saturation regime with ultra-cold atoms}

Since cold atoms have higher atomic density, one can operate in the
low saturation regime ($s\ll 1$) and still obtain strong PSR with
minimal atomic noise \cite{josse1}, when off-resonance. In the Kerr limit
($\Delta\gg\sqrt{I_{x}}\gg\gamma$), the equation of motion for the vacuum
field fluctuations is given by
\begin{equation} \label{lsr}
\frac{\partial}{\partial\bar{z}}\delta\hat{a}_y =
i\delta_0\delta\hat{a}_{y} - i\delta_{0} s (2\delta \hat{a}_{y} -
\delta \hat{a}_{y}^{\dagger})-i\frac{gNl}{c\Delta} \hat{f}_{y}.
\end{equation}
One recovers in the equation above the same terms as in the cavity
model of Ref.~\cite{josse} under the same approximations. The term in
$\delta_0\delta\hat{a}_y$ corresponds to the linear dephasing, the
second term gives the cross-Kerr squeezing term and the Langevin
noise contribution corresponding to the last term can be shown to be
proportional to $C\gamma^2/\Delta^2$, which can be small in the
off-resonant situation ($\Delta\gg\gamma$). In accordance with the
prediction of Ref.~\cite{josse} and the experimental observations of
Ref.~\cite{josse1} vacuum squeezing can be generated when
$\delta_{0}s\sim 1$ and $C\gamma^2/\Delta^2$.

\subsection{High saturation regime with thermal vapour cell}

Contrary to the situation of cold atoms, the Doppler broadening in a thermal vapour makes it impossible to work in the low saturation regime while simultaneously having low absorption or high non-linearity. It is however possible to observe strong PSR in the high saturation regime. In this regime, the atomic noise term is significantly different to that given in Eq.~(\ref{lsr}). For $I_x\gg\Delta^2$, the equation of motion is given by
\begin{eqnarray}\label{hsr}
\frac{\partial}{\partial\bar{z}}\delta \hat{a}_{y} =
\frac{i\delta_0}{2s}(\delta \hat{a}_{y} + \delta
\hat{a}_{y}^{\dagger}) - i \frac{gNl}{c\omega}(\hat{f}_{y} +
\hat{f}_{y}^{\dagger}).
\end{eqnarray}
As we have seen experimentally with the PSR measurements, the non-linear term in $\delta_{0}/(2s) = \mathcal{G}l$ can still be significant when the number of atoms are increased. However, the optical pumping processes associated with PSR now produce a lot of excess noise even in the high sideband frequency regime. The contribution of the last term in Eq.~(\ref{hsr}) can be shown to be proportional to $C\gamma^2/\omega^2\gg 1$. For our experimental parameters, the atomic noise prevents the observation of squeezing at all sideband frequencies.

\subsection{Further considerations}

We now discuss the possible discrepancies between the theoretical models
and the experiment.  Due to the complexity of the problem, many
effects have not been taken into account in the various models
discussed in this paper.

Firstly, the presence of resonance fluorescence has not been
considered in Ref.~\cite{matsko}.  In Ref.~\cite{josse}, it was
shown that PSR cannot generate squeezing in the low saturation
regime because of optical pumping processes.  We have shown in this
paper that this is also true in the high saturation regime where the
resonance fluorescence noise dominates over the cross-Kerr squeezing
term, \textit{even} at high detection frequencies. This conclusion
is in agreement with other observations \cite{boyd,agarwal,zibrov}.

Secondly, none of the models presented have included the Doppler effect.  Since we are dealing with thermal atoms, the passage of light through the atoms will give rise to a range of observed atomic detuning.  The integrated effect due to
Doppler broadening will be detrimental to the observation of squeezing.

Thirdly, the multi-level hyperfine structure of the excited states of
$^{87}$Rb have only been considered for the theoretical fits to the classical PSR results, but have not been included in any of the squeezing model. The experimental PSR data presented in this paper clearly shows that the multi-level hyperfine structure causes observable asymmetry in the PSR spectrum.  This feature cannot be explained by any of the theoretical models presented in Sec.~\ref{thy}. The multi-level theory can be expanded to include Langevin noise terms. However, a simple 4-level atom model is sufficient to demonstrate the lack of squeezing. 
The multi-level structure is also certainly less favourable to the generation of squeezing when compared with a simplified 4-level model. Different hyperfine levels will not contribute constructively towards a collective interaction that will
generate squeezing.  The added noise from these different levels will
add up significantly.  The inclusion of Doppler broadening and
multi-level effects would only result in a dominance of the atomic
noise term over the squeezing term.

Finally, the propagation of the transverse intensity profile of the
input field has been totally ignored in all models.  A full
treatment of the process should include the multi-modal analysis of
the evolution of the transverse field modes during propagation
through the vapour cell.  In the high saturation regime and for high
atomic densities, self-focussing is readily observed.  This is due to
the atom induced Kerr lens-effect on the optical field.  Thus the
centre of the field intensity distribution will undergo greater PSR
than the edges. The cross-Kerr non-linearity and the atomic
absorption used in our calculations is a result of an ``integrated''
effect of the various transverse modes.  It therefore does not model
accurately the situation of the experiment. Similar to the previous
argument, it is unlikely that the multi-modal consideration of the
process will yield better squeezing.

\subsection{Conclusion}

We have presented experimental results of PSR from two independent laboratories and have observed no squeezing. Instead we have observed excess noise in the output field spectrum at all sideband frequencies. We have modelled semi-classically the multi-level hyperfine structure of $^{87}$Rb and obtained theoretical fits to the experimental PSR data. Our multi-level modelling can predict the asymmetry in the PSR, that is due to the presence of other hyperfine excited states. We considered a quantum mechanical 4-level atomic model and showed that the squeezing term is overwhelmed by atomic noise terms in the situation of a thermal vapour. The effects of resonance fluorescence, the Doppler effect and the multi-level hyperfine structure of $^{87}$Rb all contribute to overwhelm the squeezing term. Therefore, it is expected that a full quantum mechanical treatment of a multi-level $^{87}$Rb atom will yield a result where squeezing cannot be generated. In spite of this, the 4-level atom model shows that squeezing can be generated in the situation of cold atoms where the Doppler effect is negligible. When the input field is off-resonance, the non-linearity is large but the absorption low, such that the atomic noise term does not overwhelm the squeezing term.

\begin{acknowledgments}
We would like to thank P.~Drummond, W.~P.~Bowen, J.~J.~Longdell and A.~Lvovsky for fruitful discussions and Paul Tant for technical support.  This research was funded under the Australian Research Council Centre of Excellence Programme and the European Project n$^{\circ}$FP6-511004 (COVAQIAL).
\end{acknowledgments}


\begin{thebibliography}{99}

\bibitem{mckenzie}{K.~McKenzie, D.~A.~Shaddock, D.~E.~McClelland,
B.~C.~Buchler, and P.~K.~Lam, Phys.~Rev.~Lett.  {\bf 88}, 231102
(2002).}

\bibitem{ou}{Z.~Y.~Ou, S.~F.~Pereira, H.~J.~Kimble, and K.~C.~Peng,
Phys.~Rev.~Lett.  {\bf 68}, 3663 (1992).}

\bibitem{furusawa}{A.~Furusawa, J.~L.~Sorensen, S.~L.~Braunstein,
C.~A.~Fuchs, H.~J.~Kimble, and E.~S.~Polzik, Science {\bf 282}, 706
(1998).}

\bibitem{bowen}{W.~P.~Bowen, N.~Treps, B.~C.~Buchler, R.~Schnabel,
T.~C.~Ralph, H.-A.~Bachor, T.~Symul, and P.~K.~Lam, Phys.~Rev.~A
{\bf 67}, 032302 (2003).}

\bibitem{duan}{L.-M.~Duan, M.~D.~Lukin, J.~I.~Cirac, and P.~Zoller,
Nature {\bf 414}, 413 (2001).}

\bibitem{lam}{P.~K.~Lam, T.~C.~Ralph, B.~C.~Buchler, D.~E.~McClelland,
H.-A.~Bachor, and J.~Gao, J.~Opt.~B:~Quantum~Semiclassical~Opt.  {\bf
1}, 469 (1999).}

\bibitem{sorensen}{J.~L.~Sorensen, J.~Hald, and E.~S.~Polzik,
Phys.~Rev.~Lett.  {\bf 80}, 3487 (1998).}

\bibitem{lambrecht}{A.~Lambrecht, T.~Coudreau, A.~M.~Steinberg, and
E.~Giacobino, Europhys.~Lett.  {\bf 36}, 93 (1996).}

\bibitem{josse1}{V.~Josse, A.~Dantan, L.~Vernac, A.~Bramati,
M.~Pinard, and E.~Giacobino, Phys.~Rev.~Lett.  {\bf 91}, 103601
(2003).}

\bibitem{matsko}{A.~B.~Matsko, I.~Novikova, G.~R.~Welch, D.~Budker,
D.~F.~Kimball, and S.~M.~Rochester, Phys.~Rev.~A {\bf 66}, 043815
(2002).}

\bibitem{rochester}{S.~M.~Rochester, D.~S.~Hsiung, D.~Budker,
R.~Y.~Chiao, D.~F.~Kimball, and V.~V.~Yashchuk, Phys.~Rev.~A {\bf 63},
043814 (2001).}

\bibitem{novikova}{I.~Novikova, A.~B.~Matsko, and G.~R.~Welch,
J.~Mod.~Opt.  {\bf 49}, 2565 (2002).}

\bibitem{haus}{L.~Boivin, and H.~A.~Haus, Opt.~Lett.  {\bf 21}, 146
(1996).}

\bibitem{margalit}{M.~Margalit, C.~X.~Yu, E.~P.~Ippen, and H.~A.~Haus,
Opt.~Express {\bf 2}, 72 (1998).}

\bibitem{bergman}{K.~Bergman, and H.~A.~Haus, Opt.~Lett.  {\bf 16},
663 (1991).}

\bibitem{bergman1}{K.~Bergman, C.~R.~Doerr, H.~A.~Haus, and
M.~Shirasaki, Opt.~Lett.  {\bf 18}, 643 (1993).}

\bibitem{rosenbluh}{M.~Rosenbluh and R.~M.~Shelby, Phys.~Rev.~Lett. {\bf 66}, 000153 (1991).}

\bibitem{silberhorn}{Ch.~Silberhorn, P.~K.~Lam, O.~Weiss, F.~K\"onig,
N.~Korolkova and G.~Leuchs, Phys.~Rev.~Lett.~{\bf 86}, 4267 (2001).}

\bibitem{ries}{J.~Ries, B.~Brezger, and A.~I.~Lvovsky, Phys.~Rev.~A
{\bf 68}, 025801 (2003).}

\bibitem{josse}{V.~Josse, A.~Dantan, A.~Bramati, M.~Pinard, and
E.~Giacobino, J.~Opt.~B:~Quantum~Semiclassical~Opt.  {\bf 5}, S513
(2003).}

\bibitem{boyd}{M.~Kauranen, A.~L.~Gaeta, R.~W.~Boyd, and
G.~S.~Agarwal, Phys.~Rev.~A {\bf 50}, R929 (1994).}

\bibitem{agarwal}{G.~S.~Agarwal, and R.~W.~Boyd, Phys.~Rev.~A {\bf
38}, 4019 (1988).}

\bibitem{zibrov}{A.~S..~Zibrov and I.~Novikova, e-print quant-ph/0508220.}

\end{thebibliography}
\end{document}